\documentclass[11pt]{article}
\usepackage{amssymb}
\usepackage{amstex}
\author{K.Anguige}
\title{\textbf{Isotropic Cosmological Singularities III.\\The Cauchy problem for the inhomo-\\geneous conformal Einstein-Vlasov equations.}}
\begin{document}
\maketitle
\begin{abstract}
We consider the conformal Einstein equations for  massless collisionless gas cosmologies which admit an isotropic singularity. It is shown that the Cauchy problem for these equations is well-posed with data consisting of the limiting particle distribution function at the singularity.
\end{abstract}

\section{Introduction}
In the first two papers of this series (Anguige and Tod 1998a,b, henceforth ATI, ATII) consideration was given to the Cauchy problem for the conformal Einstein equations near an isotropic singularity for two different matter models: the perfect fluid (ATI) and the collisionless gas (ATII). This problem was solved in full generality for the perfect fluid but for the collisionless gas we had to assume spatial homogeneity to make progress. In this paper we again take the collisionless gas as matter model, the aim being to extend the existence and uniqueness results of (ATII) to the inhomogeneous case. 

Recall that a spacetime~$(\tilde{M},\tilde{g}_{ab})$~has an isotropic singularity if there exists a manifold~$M\supset\tilde{M}$, a smooth Lorentz metric~$g_{ab}$~on~$M$, and a function~$Z$~defined on~$M$~such that
\begin{equation}\tilde{g}_{ab}=Z^{2}g_{ab}~~~\textrm{for}~~Z>0\end{equation}
\begin{equation}Z\rightarrow 0~~~\textrm{on}~\Sigma\end{equation}
for some spacelike hypersurface~$\Sigma$~in~$M$. Given some matter model one seeks to generate cosmologies with isotropic singularities by solving the Cauchy problem for the conformal Einstein equations, with data given for the unphysical metric~$g_{ab}$~and the matter at the singularity~$\Sigma$.

In (ATI) it was shown that with a polytropic perfect fluid as matter source the free data for the conformal field equations consist of just the 3-metric of the singularity: the second fundamental form of the singularity has to vanish and there is no data for the matter. Every choice of 3-metric then determines a unique~$\gamma-$law cosmology with an isotropic singularity. 

With a collisionless gas as source the field equations of general relativity are known as the Einstein-Vlasov (EV) equations, for the metric~$\tilde{g}_{ab}$~and the particle distribution function~$\tilde{f}$. In (ATII) it was shown that the massless EV system transforms nicely under a conformal rescaling as in (1)-(2) and that one may identify the free data for the field equations at the singularity surface~$\Sigma$. The situation is rather different from the fluid case. One is free to prescribe the limiting particle distribution function~$f^{0}$~at the singularity, subject to a single integral constraint. This function then determines the first and second fundamental forms of the singularity surface. The second fundamental form need not be zero here. 

Under the assumption of spatial homogeneity for the metric and the matter it was proved in (ATII) that each~$f^{0}$~determines a unique EV cosmology with an isotropic singularity. In the present paper we show that the same result holds in the absence of any symmetries.

With a convenient choice of conformal gauge and in harmonic coordinates, the inhomogeneous field equations may be written as a symmetric hyperbolic system of integro-differential equations with a~$\frac{1}{Z}$~forcing term, for certain carefully chosen matter and metric variables. These equations are essentially of the form studied in (Claudel and Newman 1998), but the matter integrals present in the Einstein equations prevent a direct application of the existence and uniqueness result obtained there. It is however possible to combine certain techniques used in this work with the standard method of energy estimates to obtain the desired result. We note that the semi-group theory used by Claudel and Newman is entirely avoided in our approach.

The paper is organised as follows:

In section 2 a brief review of the Einstein-Vlasov sytem and its behaviour under conformal rescaling is given.

In section 3 we collect a few results from (ATII) on the initial data for the field equations and on conformal gauge fixing.

Section 4 contains a proof of the main result, which is summarised in Theorem 4.1 at the end of the paper.

\section{The massless Einstein-Vlasov system and conformal rescaling}
A collisionless gas in GR is described by a positive function~$f(x^{a},p_{b})$~on the spacetime cotangent bundle, which represents the number of particles near the point~$x^{a}$~with 4-momentum near~$p_{b}$~(Ehlers 1971). The condition that the matter be collision-free is equivalent to requiring that~$f$~be constant along the geodesic flow, which is the vector field defined, in local coordinates, by\begin{equation}\mathcal{L}=g^{ab}p_{a}\frac{\partial}{\partial
x^{b}}-\frac{1}{2}p_{a}p_{b}\frac{\partial g^{ab}}{\partial
x^{c}}\frac{\partial}{\partial p_{c}}\end{equation}
The statement~$\mathcal{L}f=0$~is known as the Vlasov equation.

For massless particles the function~$f$~is supported on the seven-dimensional submanifold of the cotangent bundle given by the equation~$g^{ab}p_{a}p_{b}=0$. The stress-energy-momentum tensor due to massless particles is given by
\begin{equation}T_{ab}=\int_{\mathbb{R}^{3}}fp_{a}p_{b}\frac{(-g)^{-1/2}}{p^{0}}d^{3}p\end{equation}
with the positive quantity~$p^{0}$~being determined by the relation~$g^{ab}p_{a}p_{b}=0$. Note that the massless condition implies~$T^{a}_{~a}=0$.
 
If the Vlasov equation holds then this stress tensor is divergence-free independently of the Einstein equations being satisfied.
 
The coupled Einstein-Vlasov equations, for the
metric~$g_{ab}$~and the particle distribution~$f$~are
\begin{equation}G_{ab}=8\pi\int_{\mathbb{R}^{3}}fp_{a}p_{b}\frac{(-g)^{-1/2}}{p^{0}}d^{3}p\end{equation}
\begin{equation}\mathcal{L}_{g}(f)=0\end{equation}

Suppose now that we have a physical spacetime~$(\tilde{M},\tilde{g}_{ab})$~and a massless particle distribution function~$\tilde{f}$ so that
~$\tilde{g}_{ab}$~and~$\tilde{f}$~satisfy tilded versions of
(5)-(6). Suppose also that we have an unphysical spacetime~$(M,g_{ab})$~defined by
\begin{equation}\tilde{g}_{ab}=~Z^{2}g_{ab}\end{equation}
for some conformal factor~$Z$~on~$M\supset\tilde{M}$.

As discussed in (AT) it is a consequence of the conformal invariance of null geodesics that the massless Vlasov equation is also conformally invariant. This is to say that if we simply write~$f\equiv\tilde{f}$~then the Vlasov equation for~$\tilde{f}$~in~$\tilde{M}$~implies the Vlasov equation for~$f$~in~$M$:
\begin{equation}\mathcal{L}_{g}f=0\end{equation}
If we now use the unphysical metric~$g_{ab}$~to define an unphysical stress tensor via
\begin{equation}T_{ab}=\int~fp_{a}p_{b}\frac{(-g)^{-1/2}}{p^{0}}d^{3}p\end{equation}
then
\begin{equation}\tilde{T}_{ab}=\frac{1}{Z^{2}}T_{ab}\end{equation}
and one may now use the unphysical Vlasov equation (8) to show that\linebreak[4]$\nabla^{a}T_{ab}=0$~in~$M$, just as~$\tilde{\nabla}^{a}\tilde{T}_{ab}=0$~in~$\tilde{M}$ .

The conformal EV equations for~$g_{ab}$~and~$f$~are now
\begin{displaymath}R_{ab}=2\nabla_{a}\nabla_{b}\log Z-2\nabla_{a}\log Z\nabla_{b}\log Z\end{displaymath}
\begin{equation}+g_{ab}(\square
\log Z+2\nabla_{c}\log Z\nabla^{c}\log Z)+\frac{8\pi}{Z^{2}}T_{ab}\end{equation}
and
\begin{equation}\mathcal{L}_{g}(f)=0\end{equation}
From now on we assume that a rescaling of the spacetime metric as in (1)-(2) exists, and our aim will be to solve equations (11)-(12) for~$f$~and~$g_{ab}$~in the neighbourhood of the singularity surface~$\Sigma$. 
\\
In (ATII) it was noted that the massless condition on the matter implies that the conformal factor~$Z$~must remain smooth at~$\Sigma$. In what follows we will require~$Z$~to be a good coordinate in~$M$, and we thus make the following extra assumption:
\\\\\textbf{Assumption~2.1}~\textit{The conformal
factor}~$Z$~\textit{is such that}~$\nabla_{a}Z\neq 0$~\textit{at}~$\Sigma$.\\\\ 
\section{Initial data and conformal gauge fixing}
The singular nature of the conformal field equations (11),(12) imposes constraints on the data at an isotropic singularity~$\Sigma$~,which are rather different from the usual constraints on a regular hypersurface. To say what these are we first make the following definitions:\\
Let~$h^{0}_{ij}$~be the 3-metric of the singularity surface~$\Sigma$~and
let~$K^{0}_{ij}$~be the second fundamental form of~$\Sigma$~in~$M$.~$f^{0}$~will stand for the orthogonal projection of~$f$~onto~$\Sigma$~:~$f^{0}(x^{i},p_{b}^{~\perp})=f(0,x^{i},p_{b})$~and~$V^{0}=(g^{ab}\nabla_{a}Z\nabla_{b}Z)^{1/2}$~, as evaluated at~$\Sigma$.\\\\
The constraints then read as follows: 
\begin{equation}\int f^{0}(x^{i},p_{j})p_{k}~d^{3}p=0\end{equation}
\begin{equation}(V^{0})^{2}h^{0}_{ij}=\frac{8\pi}{\sqrt{\textrm{det}h^{0}}}\int\frac{f^{0}p_{i}p_{j}}{((h^{0})^{mn}p_{m}p_{n})^{1/2}}~d^{3}p\end{equation}
\begin{equation}\Big(2\delta^{i}_{~k}\delta^{j}_{~l}-\chi^{ij}_{~~kl}\Big)(K^{0})^{kl}=D^{k}(\chi^{ij}_{~~k})+\frac{1}{2}(K^{0})(h^{0})^{ij}\end{equation}
where~$K^{0}=(h^{0})^{ij}K^{0}_{ij}$~,
\begin{equation}\chi_{ijkl}\equiv\frac{4\pi}{(V^{0})^{2}\sqrt{h^{0}}}\int\frac{f^{0}p_{i}p_{j}p_{k}p_{l}}{((h^{0})^{mn}p_{m}p_{n})^{3/2}}~d^{3}p\end{equation}
\begin{equation}\chi_{ijk}\equiv
-\frac{4\pi}{(V^{0})^{2}\sqrt{h^{0}}}\int\frac{f^{0}p_{i}p_{j}p_{k}}{((h^{0})^{mn}p_{m}p_{n})}~d^{3}p\end{equation}
and~$D_{i}$~is the covariant derivative operator associated with~$h^{0}_{ij}$.\\
Now (13) is just an integral constraint on~$f^{0}$~and from (ATII) we have the following theorem:

\textbf{Theorem 3.1} Let~$f^{0}(x^{i},p_{j})$~be a smooth, positive function on~$U\times \mathbb{R}^{3}$, where~$U$~is an open subset of~$\mathbb{R}^{3}$. Suppose that for each~$x\in U$
\begin{enumerate}
\item~~$f^{0}$~is compactly supported in~$p$.
\item~~$f^{0}$~is supported outside some open ball containing~$p=0$.
\item~~$f^{0}$~is not identically zero in~$p$.
\end{enumerate}
Then, given a smooth strictly positive function~$V^{0}$~on~$U$, there exists a unique, smooth, positive definite 3-metric~$h^{0}_{ij}$~on~$U$~satisfying (14) and a unique smooth trace-free symmetric tensor~$K^{0}_{ij}$~satisfying (15)\\\\
Also from (ATII) we have that the freedom in the choice of unphysical metric~$g_{ab}$~may be used to obtain the following:

\textbf{Theorem 3.2} If~$(\tilde{M},\tilde{g}_{ab})$~is a solution of the
massless Einstein-Vlasov equations, with an isotropic singularity, and
Assumption 2.1 holds, then the conformal factor~$Z$~may be chosen so
that~$V^{0}=1,~K^{0}=0$, and~$Z$~is a harmonic function
in~$M$.\\\\
We will henceforth refer to the gauge choice of Theorem 3.2 as the \linebreak\textit{harmonic}~gauge.\\
In summary the initial data set for the conformal EV equations at~$Z=0$, in the harmonic gauge, is as follows:
\\
\begin{itemize}
\item~$f^{0}$~is free data, subject only
to the integral constraint (13).
\item The initial
3-metric~$h^{0}_{ij}$~is determined by ~$f^{0}$~via~(14) \linebreak with ~$V^{0}\equiv 1$.
\item The initial second fundamental form~$K^{0}_{ij}$~is determined by
~$f^{0}$~via (15) with~$V^{0}\equiv 1$,~$K^{0}\equiv 0$
\end{itemize}
\section{The Cauchy problem for the conformal EV equations near~$\Sigma$}
In this section we show that the harmonic gauge conformal EV equations may be solved by a combination of the standard method of energy estimates with certain of the techniques used in the proof of the Newman-Claudel theorem. This theorem cannot be applied directly because of the presence of matter integrals in the field equations.

\subsection{The reduced Einstein equations as a symmetric hyperbolic system}
The first step in solving the field equations is to impose the harmonic coordinate condition which leads to a system of reduced equations.

Given a coordinate system ~$x^{\mu}$~ we make the following definition
\begin{equation}R^{H}_{\mu\nu}\equiv
R_{\mu\nu}+g_{\alpha(\mu}\partial_{\nu)}H^{\alpha}\end{equation}
where ~$H^{\alpha}=\square x^{\alpha}$.
\\
The reduced conformal EV equations are then defined to be
\begin{equation}R^{H}_{\mu\nu}-\frac{2}{Z}\nabla_{\mu}\nabla_{\nu}Z+\frac{4}{Z^{2}}\nabla_{\mu}Z\nabla_{\nu}Z-\frac{V^{2}}{Z^{2}}g_{\mu\nu}=\frac{1}{Z^{2}}T_{\mu\nu}\end{equation}
\begin{equation}\mathcal{L}_{g}f=0\end{equation}
If we
put~$x^{0}=Z$~and~$h_{\alpha\beta\gamma}=g_{\alpha\beta,\gamma}$, then
equations (19)-(20) can be decomposed as
\begin{equation}\partial_{z}g_{\alpha\beta}=h_{\alpha\beta 0}\end{equation}
\begin{equation}-g^{ij}\partial_{z}h_{\alpha\beta
j}=-g^{ij}\partial_{j}h_{\alpha\beta 0}\end{equation}
\begin{equation}\partial_{z}g^{\alpha\beta}=-g^{\alpha\mu}g^{\beta\nu}h_{\mu\nu
0}\end{equation}
\begin{displaymath}g^{00}\partial_{z}h_{\alpha\beta
0}=-g^{ij}\partial_{i}h_{\alpha\beta
j}-2g^{0i}\partial_{i}h_{\alpha\beta 0}+F_{\alpha\beta}(g,g^{-1},h)\end{displaymath}
\begin{equation}+\frac{1}{Z}\{Z_{\alpha\beta}+2g^{0\gamma}(h_{\alpha\gamma\beta}+h_{\beta\gamma\alpha}-h_{\alpha\beta\gamma})\}\end{equation}
\begin{equation}p^{0}\frac{\partial f}{\partial Z}+p^{i}\frac{\partial
f}{\partial
x^{i}}-\frac{1}{2}(\partial_{i}g^{\alpha\beta})p_{\alpha}p_{\beta}\frac{\partial
f}{\partial p_{i}}=0\end{equation}
where
\begin{equation}Z_{\mu\nu}\equiv\frac{2}{Z}\{4\nabla_{\mu}Z\nabla_{\nu}Z-g^{00}g_{\mu\nu}-T_{\mu\nu}\}\end{equation}
and ~$F_{\alpha\beta}$~is polynomial in its arguments.
\\
It follows that the reduced Einstein equations (21)-(24) are, for fixed ~$f$, in symmetric hyperbolic form.

\subsection{Initial data for the reduced equations}
We would like to solve the conformal EV equations in
the harmonic gauge. So, as initial data for the reduced equations first choose a positive smooth function~$f^{0}$~on~$T^{\star}\Sigma$~satisfying the constraint (13). We also require~$f^{0}(x^{i},\cdot)$\linebreak[4]to have fixed compact support and to be supported away from a fixed neighbourhood of the origin for~$x^{i}$~in a compact set. Then put~$(g^{0})^{00}=(V^{0})^{2}=1$~and choose~$g^{0}_{0i}=0$, so that the~$x^{i}$~are `initially comoving'. We may now use Theorem 3.1 to determine~$g^{0}_{ij}$~and~$h^{0}_{ij0}=2K^{0}_{ij}$~from (14) and (15) with ~$K^{0}$~set to zero.
\\
From (19) one must have that~$\square Z=0$~initially. But by
definition
\begin{equation}\square Z=\partial_{\alpha}g^{\alpha
0}+\frac{1}{2}g^{\alpha
0}g^{\rho\sigma}\partial_{\alpha}g_{\rho\sigma}\end{equation}
so that ~$h_{000}=0$~initially.
\\
Equation (24) implies
\begin{equation}Z^{0}_{\alpha\beta}=2\{h^{0}_{\alpha\beta
0}-h^{0}_{\alpha 0\beta}-h^{0}_{\beta 0\alpha}\}\end{equation}
so that~$Z^{0}_{ij}=2h^{0}_{ij0},~Z^{0}_{00}=Z^{0}_{0i}=0$.
\\
In order to recover the full harmonic gauge equations from the reduced equations we
will have to show that~$H^{\alpha}\equiv\square
x^{\alpha}=0$~in~$M$. We therefore pick~$h^{0}_{i00}=0$~and choose
spatial coordinates~$x^{i}$~satisfying the following condition on~$\Sigma$:
\begin{equation}g^{mn}\Gamma^{i}_{mn}=0\end{equation}
where~$\Gamma^{i}_{jk}$~are the Christoffel symbols
of~$g^{0}_{ij}$. These choices ensure that~$\square x^{i}=0$~on~$\Sigma$.
\subsection{Recovering the harmonic gauge equations}
We now use the contracted Bianchi identities to show that if ~$(f,g_{\alpha\beta})$~is a solution of the
reduced equations (19)-(20) with data as prescribed in 4.2,
then~$\square Z=\square x^{i}=0$~and~$(f,g_{\alpha\beta})$~solves the
conformal EV equations (11)-(12).\\
First write
\begin{equation}S_{\mu\nu}\equiv
\frac{1}{Z^{2}}T_{\mu\nu}+\frac{2}{Z}\nabla_{\mu}\nabla_{\nu}Z+\frac{V^{2}}{Z^{2}}g_{\mu\nu}-\frac{4}{Z^{2}}\nabla_{\mu}Z\nabla_{\nu}Z\end{equation}
We know from (20) that~$\nabla^{\mu}T_{\mu\nu}=0$~and this leads to
\begin{equation}\nabla^{\mu}S_{\mu\nu}=\frac{2}{Z}(\nabla^{\alpha}Z)(R_{\alpha\nu}-S_{\alpha\nu})+\frac{2}{Z}\partial_{\nu}H^{0}-\frac{4}{Z^{2}}(\partial_{\nu}Z)H^{0}\end{equation}
and now (19) gives
\begin{equation}\nabla^{\mu}R^{H}_{\mu\nu}=\frac{2}{Z}(\nabla^{\alpha}Z)(-g_{\beta(\alpha}\partial_{\nu)}H^{\beta})+\frac{2}{Z}\partial_{\nu}H^{0}-\frac{4}{Z^{2}}(\partial_{\nu}Z)H^{0}\end{equation}
From (19) one has~$R^{H}=\frac{2}{Z}H^{0}$~, so that
\begin{equation}R=\frac{2}{Z}H^{0}-g^{\alpha\beta}g_{\gamma(\alpha}\partial_{\beta
 )}H^{\gamma}\end{equation}
The contracted Bianchi identities,~$\nabla^{\mu}G_{\mu\nu}=0$,~now imply
\begin{displaymath}0=\frac{2}{Z}(\nabla^{\alpha}Z)(-g_{\beta(\alpha}\partial_{\nu)}H^{\beta})+\frac{2}{Z}\partial_{\nu}H^{0}-\frac{4}{Z^{2}}(\partial_{\nu}Z)H^{0}\end{displaymath}
\begin{equation}-\nabla^{\alpha}(g_{\beta(\alpha}\partial_{\nu)}H^{\beta})-\frac{1}{2}\nabla_{\nu}\Big(\frac{2}{Z}H^{0}-g^{\alpha\beta}g_{\gamma(\alpha}\partial_{\beta)}H^{\gamma}\Big)+F_{\nu}(g,g^{-1},h,H,\partial
H)\end{equation}
where~$F_{\nu}$~is polynomial in its arguments and homogeneous
in~$H,~\partial H$.
\\
Now multiply (34) by~$g^{\nu\lambda}$~to get
\begin{displaymath}0=-\frac{1}{2}g^{\rho\mu}\partial^{2}_{\rho\mu}H^{\lambda}-\frac{2}{Z}(\nabla^{\alpha}Z)g^{\nu\lambda}g_{\beta(\alpha}\partial_{\nu)}H^{\beta}\end{displaymath}
\begin{equation}+\frac{1}{Z}g^{\nu\lambda}\partial_{\nu}H^{0}-\frac{3}{Z}g^{\nu\lambda}(\partial_{\nu}Z)P^{0}+\hat{F}^{\lambda}\end{equation}
where~$P^{0}\equiv\frac{1}{Z}H^{0}$.
\\
From (35) one gets the following at~$Z=0$:
\begin{equation}-2(\nabla^{\alpha}Z)g^{\nu\lambda}g_{\beta(\alpha}\partial_{\nu)}H^{\beta}+g^{\nu\lambda}\partial_{\nu}H^{0}-3g^{\nu\lambda}(\partial_{\nu}Z)P^{0}=0\end{equation}
Now~$P^{0}(0)=\partial_{z}H^{0}(0)$, so (36) implies
that~$P^{0}=\partial_{z}H^{0}=0$~at$~Z=0$. Equation (36) also
implies that~$\partial_{z}H^{i}=0$~at~$Z=0$.\\
From the definition of ~$P^{0}$~it follows that
\begin{equation}\frac{\partial P^{0}}{\partial
Z}=\frac{1}{Z}\Big(-P^{0}+\frac{\partial H^{0}}{\partial
Z}\Big)\end{equation}
If we put~$h_{\alpha}^{\beta}\equiv \partial_{\alpha}H^{\beta}$~then
(35) and (37) can be written in first order form as
\begin{equation}-g^{ij}\frac{\partial h_{j}^{\alpha}}{\partial
Z}=-g^{ij}\frac{\partial h_{0}^{\alpha}}{\partial x^{j}}\end{equation}
\begin{equation}\frac{\partial H^{\alpha}}{\partial
Z}=h_{0}^{\alpha}\end{equation}
\begin{equation}\frac{\partial P^{0}}{\partial
Z}=-\frac{1}{Z}P^{0}+\frac{1}{Z}h_{0}^{0}\end{equation}
\begin{displaymath}(g^{00})\frac{\partial h_{0}^{\lambda}}{\partial
Z}=2g^{0i}\partial_{i}h_{0}^{\lambda}-g^{ij}\partial_{i}h_{j}^{\lambda}\end{displaymath}
\begin{equation}-\frac{2}{Z}\{g^{0\alpha}h_{\alpha}^{\lambda}+3g^{0\lambda}P^{0}\}+\tilde{F}^{\lambda}\end{equation}
The system (38)-(41) can be written in an obvious way as:
\begin{equation}a^{0}(u)\frac{\partial u}{\partial
Z}=a^{i}(u)\frac{\partial u}{\partial
x^{i}}+b(u)u+\frac{1}{Z}c(u)u\end{equation}
with~$a^{0}$~+ve definite and~$a^{\alpha}$~symmetric.
\\
The eigenvalues ~$\lambda$~of~$((a^{0})^{-1}c)(0)$~satisfy either
\begin{equation}\lambda=0\end{equation}
or
\begin{equation}\lambda P^{0}=-P^{0}+h_{0}^{0}\end{equation}
\begin{equation}\lambda h_{0}^{\alpha}=-2g^{0\beta}(0)h_{\beta}^{\alpha}-6g^{0\alpha}(0)P^{0}\end{equation}
Putting~$\alpha=i$~in (45) gives
\begin{displaymath}\lambda h_{0}^{i}=-2h_{0}^{i}\end{displaymath}
while putting~$\alpha=0$~in (45) gives
\begin{equation}\lambda h_{0}^{0}=-2h_{0}^{0}-6P^{0}\end{equation}
So if ~$\lambda\neq~0,-2$~then
\begin{equation}(2+\lambda)(1+\lambda)+6=0\end{equation}
Thus~$(a^{0})^{-1}c(0)$~has no positive integer eigenvalues. Now, as remarked in section 4.2, the choice of initial data for the reduced equations forces the quantities~$\square x^{\alpha}$~to vanish at~$Z=0$. Therefore (42) has the trivial solution~$u\equiv 0$~and by Theorem A.1 (see Appendix) this is the only smooth solution with the given data. It follows that any smooth solution of the reduced equations (19)-(20), with data as in 4.2, is a solution of the harmonic gauge EV equations .
\subsection{Solving the reduced field equations}
We aim to solve the reduced EV equations on a manifold of the form~$M\times [0,T]$, with~$M$~a paracompact 3-manifold. We will use the method of energy estimates to obtain local (in space and time) solutions of the field equations, which can then be patched together using the finite domain of dependence to obtain 
a solution in a neighbourhood of the singularity~$\Sigma=M\times \{0\}$. 
\subsubsection{Localisation}
First choose a locally finite cover of~$\Sigma$~by relatively compact harmonic coordinate charts~$O_{\alpha}$. Now construct a second cover~$\{O'_{\alpha}\}$~in such a way that~$\bar{O}'_{\alpha}\subset O_{\alpha}$.\\
Restrict attention to a single chart~$O_{\alpha}$~and consider the reduced equations (19)-(20) on~$O_{\alpha}\times [0,T]$. In order to obtain energy estimates it is desirable to have data in~$C_{0}^{\infty}(O_{\alpha})$. We thus cut off~$f^{0}$~to zero smoothly, along with~$g_{ij}+\delta_{ij}$. We now work with~$\bar{g}_{\alpha\beta}=g_{\alpha\beta}-\eta_{\alpha\beta}$~as field variable ($\eta_{\alpha\beta}=\textrm{diag}(1,-1,-1,-1)$). Obtain~$K^{0}_{ij}$~from the new~$f^{0},~\bar{g}_{\alpha\beta}$~via equation (15). This cutting-off procedure is done in such a way that the original data is maintained on~$O_{\alpha}''\supset\bar{O'}_{\alpha}$~and is zero outside~$O_{\alpha}'''\supset\bar{O}_{\alpha}''$.

Now note that the singular field equation (26) will no longer be consistent at~$Z=0$~with the localised~$C_{0}^{\infty}$~data. To remedy this we introduce a function~$F(x^{i})\in C_{0}^{\infty}(O_{\alpha})$~such that~$0\leq F(x^{i})\leq 1$~and
\begin{equation}F(x)=\left\{\begin{array}{ll}0&~~~x\in O_{\alpha}\setminus O_{\alpha}''\\1&~~~x\in O_{\alpha}'''':~O_{\alpha}''\supset O_{\alpha}''''\supset\bar{O}_{\alpha}'\end{array}\right.\end{equation}
We use the function~$F$~to obtain the following localised modification of (26)-(27):
\begin{equation}R^{H}_{\mu\nu}=F(x^{i})\left\{\frac{2}{Z}\nabla_{\mu}\nabla_{\nu}Z+\frac{1}{Z^{2}}(T_{\mu\nu}+g^{00}g_{\mu\nu}-4\nabla_{\mu}Z\nabla_{\nu}Z)\right\}\end{equation}
\begin{equation}L_{g}f=0\end{equation}
We will solve (49)-(50) on~$O_{\alpha}\times [0,T_{\alpha}]$~and then use the finite domain  of dependence to argue that this gives rise to a solution of conformal Einstein-Vlasov on~$O_{\alpha}'\times [0,T_{\alpha}']$, determined by the right data. 

\subsubsection{Approximate solutions of the field equations}
An important step in solving the reduced field equations is, as in the proof of the Newman-Claudel theorem, to write them in such a way that they are manifestly formally solvable. One then obtains approximate finite Taylor series solutions which solve the equations up to~$O(Z^{q})$,~$q$~arbitrary. We note that the hyperbolicity of the field equations is not used at this stage.
\\\\
First define
\begin{equation}Z_{\mu\nu}=\frac{2}{Z}\left\{4\nabla_{\mu}Z\nabla_{\nu}Z-g^{00}g_{\mu\nu}-T_{\mu\nu}-(4\nabla_{\mu}Z\nabla_{\nu}Z-g^{00}g_{\mu\nu}-T_{\mu\nu})(0)\right\}\end{equation}
Then (49) is just
\begin{equation}R^{H}_{\mu\nu}=F(x^{i})\left\{\frac{2}{Z}\nabla_{\mu}\nabla_{\nu}Z+\frac{1}{2Z}Z_{\mu\nu}\right\}\end{equation}
(F was chosen to be zero wherever~$(4\nabla_{\mu}Z\nabla_{\nu}Z-g^{00}g_{\mu\nu}-T_{\mu\nu})(0)\neq 0$)\\
Define also
\begin{equation}\bar{Z}_{\alpha\beta}=Z_{\alpha\beta}-Z_{\alpha\beta}(0)~~~~\bar{h}_{\alpha\beta\gamma}=h_{\alpha\beta\gamma}-h_{\alpha\beta\gamma}(0)\end{equation}
In terms of these variables, equations (49)-(50) become
\begin{equation}-g^{ij}\partial_{z}\bar{h}_{\alpha\beta
j}=-g^{ij}\partial_{j}(\bar{h}_{\alpha\beta 0}+h_{\alpha\beta
0}(0))\end{equation}
\begin{equation}\partial_{z}\bar{g}_{\alpha\beta}=\bar{h}_{\alpha\beta
0}+h_{\alpha\beta 0}(0)\end{equation}
\begin{equation}\partial_{z}\bar{g}^{\alpha\beta}=-g^{\beta\nu}g^{\alpha\mu}(\bar{h}_{\mu\nu
0}+h_{\mu\nu 0})\end{equation}
\begin{displaymath}g^{00}\partial_{z}\bar{h}_{\alpha\beta
0}=-g^{ij}\partial_{i}(\bar{h}_{\alpha\beta j}+h_{\alpha\beta
j}(0))-2g^{0i}\partial_{i}(\bar{h}_{\alpha\beta 0}+h_{\alpha\beta
0}(0))\end{displaymath}
\begin{displaymath}+2\widehat{F}_{\alpha\beta}+2F(x^{i})(F_{1})^{\gamma}(\bar{h}_{\alpha\gamma\beta}+\bar{h}_{\beta\gamma\alpha}-\bar{h}_{\alpha\beta\gamma})\end{displaymath}
\begin{equation}+\frac{F(x^{i})}{Z}\{\bar{Z}_{\alpha\beta}+(H_{1})^{\gamma}(0)(\bar{h}_{\alpha\gamma\beta}+\bar{h}_{\beta\gamma\alpha}-\bar{h}_{\alpha\beta\gamma})\}\end{equation}
and
\begin{equation}L_{g}f=0\end{equation}
where
\begin{equation}(F_{1})^{\gamma}(Z)\equiv
\int_{0}^{1}((H_{1})^{\gamma})'(SZ)~dS~,~~~(H_{1})^{\gamma}\equiv g^{0\gamma}\end{equation}
The prime denotes differentiation with respect to time, and~$F_{1}$~ can be computed from (53) and (56).\\
Meanwhile we calculate the following evolution for~$\bar{Z}_{\mu\nu}$
\begin{displaymath}\partial_{z}\bar{Z}_{\mu\nu}=-F_{2}(\bar{h}_{\mu\nu
0}+h_{\mu\nu
0}(0))+(F_{3})^{\alpha\beta}_{~~~\mu\nu}(\bar{h}_{\alpha\beta
0}+h_{\alpha\beta 0}(0))\end{displaymath}
\begin{displaymath}+(F_{4})^{\alpha\beta}_{~~~\mu\nu}(\bar{h}_{\alpha\beta
0}+h_{\alpha\beta
0}(0))+(F_{5})_{\mu\nu}+(F_{6})^{\alpha\beta}_{~~~\mu\nu}(\bar{h}_{\alpha\beta
i}+h_{\alpha\beta i})\end{displaymath}
\begin{displaymath}-(F_{7})^{\alpha\beta}_{~~~\mu\nu}(\bar{h}_{\alpha\beta
0}+h_{\alpha\beta 0})\end{displaymath}
\begin{displaymath}+\frac{1}{Z}\Big\{-\bar{Z}_{\mu\nu}-H_{2}(0)\bar{h}_{\mu\nu
0}+(H_{3})^{\alpha\beta}_{~~~\mu\nu}(0)\bar{h}_{\alpha\beta
0}\end{displaymath}
\begin{equation}+(H_{4})^{\alpha\beta}_{~~~\mu\nu}(0)\bar{h}_{\alpha\beta
0}+(H_{6})^{\alpha\beta}_{~~~\mu\nu}(0)\bar{h}_{\alpha\beta
i}-(H_{7})^{\alpha\beta}_{~~~\mu\nu}(0)\bar{h}_{\alpha\beta
0}\Big\}\end{equation}
where ~$F_{i}$~is to~$H_{i}$~as ~$F_{1}$~is to~$H_{1}$~and
\begin{equation}H_{2}\equiv 2g^{00}\end{equation}
\begin{equation}(H_{3})^{\alpha\beta}_{~~~\mu\nu}\equiv
2g_{\mu\nu}g^{0\alpha}g^{0\beta}\end{equation}
\begin{equation}(H_{4})^{\alpha\beta}_{~~~\mu\nu}\equiv
(\textrm{det}g)^{-1/2}g^{\alpha\beta}\int\frac{fp_{\mu}p_{\nu}}{p^{0}}~d^{3}p\end{equation}
\begin{equation}(H_{5})_{\mu\nu}\equiv
2(\textrm{det}g)^{-1/2}\int\frac{p_{\mu}p_{\nu}p^{i}}{(p^{0})^{2}}\frac{\partial
f}{\partial x^{i}}~d^{3}p\end{equation}
\begin{equation}(H_{6})^{\alpha\beta}_{~~~\mu\nu}\equiv
(\textrm{det}g)^{-1/2}\int\frac{p_{\mu}p_{\nu}p^{\alpha}p^{\beta}}{(p^{0})^{2}}\frac{\partial
f}{\partial p_{i}}~d^{3}p\end{equation}
\begin{equation}(H_{7})^{\alpha\beta}_{~~~\mu\nu}\equiv (\textrm{det}g)^{-1/2}\int\frac{f}{(p^{0})^{2}}\Big(p^{\alpha}p^{\beta}((p_{0})^{-1}p_{\mu}p_{\nu}+p_{\mu}\delta^{0}_{\nu}+p_{\nu}\delta^{0}_{\mu})\Big)~d^{3}p\end{equation}
The~$F_{i}$~are regular terms, and can be calculated from (54),
(55), (58).\\
The whole system of equations (54)-(58), (60) can be written in the form
\begin{equation}A^{0}(u)\partial_{z}u=A^{i}(u)\partial_{i}u+B(u)u+\frac{1}{Z}C(x^{i},u)u\end{equation}
(i=1,2\ldots 6) for some matrices~$A^{\alpha}, B, C$~with~$A^{0}$~positive definite. The vector~$u$~stands for~$(f,\bar{g}_{\alpha\beta},\bar{g}^{\alpha\beta},\bar{h}_{\alpha\beta\gamma},\bar{Z}_{\alpha\beta}).$\\\\
~~~\textbf{Lemma 4.1}~~~The matrix~$C(x^{i},u)$~in (74) satisfies~$C(x^{i},u)u^{0}=0~\forall u$~and\linebreak[4]$(A^{0})^{-1}C(x^{i},u^{0})$~has no positive integer eigenvalues.

\textit{Proof}~Since~$\bar{Z}_{\alpha\beta}(0)=\bar{h}_{\alpha\beta\gamma}(0)=0$~the first part follows. Now note that the eigenvalues of~$((A^{0})^{-1}C)(0)$~satisfy either~$\lambda=0$~or
\begin{equation}\lambda h_{\alpha\beta 0}=F(x^{i})\left\{Z_{\alpha\beta}+\{h_{\alpha
0\beta}+h_{\beta 0\alpha}-h_{\alpha\beta 0}\}\right\}\end{equation}
\begin{displaymath}\lambda
Z_{\alpha\beta}=-Z_{\alpha\beta}+2\Bigg\{-h_{\alpha\beta0}+g_{\alpha\beta}(0)g^{0\mu}(0)g^{0\nu}(0)h_{\mu\nu
0}\end{displaymath}
\begin{displaymath}+\frac{1}{2}(\textrm{det}g(0))^{-1/2}g^{\mu\nu}(0)h_{\mu\nu
0}\int\frac{f^{0}p_{\alpha}p_{\beta}}{p^{0}}~d^{3}p\end{displaymath}
\begin{equation}-\frac{1}{2}(\textrm{det}g(0))^{-1/2}h_{\mu\nu
0}\int\frac{f^{0}p^{\mu}p^{\nu}}{(p^{0})^{2}}\{(p_{\alpha}p_{\beta}/p_{0})+p_{\alpha}\delta^{0}_{\beta}+p_{\beta}\delta^{0}_{\beta}\}~d^{3}p\Bigg\}\end{equation}
where~$p_{0}=p^{0}=(-g^{ij}(0)p_{i}p_{j})^{1/2}$~and~$h_{\alpha\beta i}=0$.\\
Now if~$F(x^{i})=0$~then either~$\lambda=0$~or~$h_{\alpha\beta 0}=0$. If~$\lambda\neq 0$~then from (68)-(69) we must have~$\lambda=-1$. It is therefore enough to consider those~$x$~at which~$F(x)\neq 0$. At such~$x$~the relation between~$f^{0}$~and~$g_{ij}(0)$~is given by (14).\\
Equations (68)-(69) can be expanded as follows:
\begin{equation}\lambda h_{000}=F(Z_{00}+2h_{000})\end{equation}
\begin{equation}\lambda
Z_{00}=-Z_{00}-6h_{000}+6g^{ij}(0)h_{ij0}\end{equation}
\begin{equation}\lambda h_{i00}=F(Z_{i0})\end{equation}
\begin{equation}\lambda
Z_{i0}=-Z_{i0}-2h_{i00}-2(\textrm{det}g(0))^{-1/2}h_{jk0}\int\frac{f^{0}}{(p_{0})^{2}}p^{j}p^{k}p_{i}~d^{3}p\end{equation}
\begin{equation}\lambda h_{ij0}=F(Z_{ij}-2h_{ij0})\end{equation}
\begin{displaymath}\lambda
Z_{ij}=-Z_{ij}-2h_{ij0}+2g_{ij}(0)h_{000}-g_{ij}(0)(h_{mn0}g^{mn}(0))\end{displaymath}
\begin{equation}-h_{mn0}\Big(\chi^{mn}_{~~~ij}\Big)\end{equation}
where
\begin{equation}\chi^{mn}_{~~~ij}=(\textrm{det}g(0))^{-1/2}\int\frac{f^{0}}{(p_{0})^{3}}p^{m}p^{n}p_{i}p_{j}~d^{3}p\end{equation}
Taking the trace of (74)-(75) leads to
\begin{equation}(g^{mn}(0)h_{mn0})((\lambda+1)(\lambda+2)+6F)=6Fh_{000}\end{equation}
while (70)-(71) give
\begin{equation}((\lambda+1)(\lambda-2F)+6F)h_{000}=6Fg^{mn}(0)h_{mn0}\end{equation}
and hence
\begin{equation}(g^{mn}(0)h_{mn0})(((\lambda+1)(\lambda-2F)+6F)((\lambda+1)(\lambda+2)+6F)-36F^{2})=0\end{equation}
Now if~$\lambda$~were a positive integer then the polynomial in~$F$~and~$\lambda$~on the left hand side of (79) would be greater than or equal to~$12(1+2F(1-F))$. Since~$0\leq F\leq 1$~it follows that if~$g^{mn}(0)h_{mn0}\neq 0$~then~$\lambda$~is not a positive integer.\\
Suppose now that~$g^{mn}(0)h_{mn0}=0$. Equations (74)-(75) then
become
\begin{equation}\lambda h_{ij0}=F(Z_{ij}-2h_{ij0})\end{equation}
\begin{equation}\lambda
Z_{ij}=-Z_{ij}-2h_{ij0}+2g_{ij}(0)h_{000}-h_{mn0}\Big(\chi^{mn}_{~~~ij}\Big)\end{equation}
and (70)-(71) become
\begin{equation}\lambda h_{000}=F(Z_{00}+2h_{000})\end{equation}
\begin{equation}\lambda Z_{00}=-Z_{00}-6h_{000}\end{equation}
which imply
\begin{equation}((\lambda+1)(\lambda-2)+6F)h_{000}=0\end{equation}
Suppose~$(\lambda+1))(\lambda-2)+6F\neq 0$, so that~$h_{000}=0$. Then (74)-(75)
imply
\begin{equation}((\lambda+1)(\lambda+2)+2F)h_{ij0}=-Fh_{mn0}\chi^{mn}_{~~~ij}\end{equation}
But clearly the eigenvalues of~$\chi$~are positive and hence~$\lambda$~cannot be a positive integer.\\
Suppose finally that~$h_{mn0}=0$. Then (72)-(73) are
\begin{equation}\lambda h_{i00}=F(Z_{i0})\end{equation}
\begin{equation}\lambda Z_{i0}=-Z_{i0}-2h_{i00}\end{equation}
which imply
\begin{equation}(\lambda^{2}+\lambda+2F)h_{i00}=0\end{equation}
and we conclude from all this that~$((A^{0})^{-1}C)(0)$~has no positive
integer eigenvalues.

Lemma 4.1 now allows us to construct approximate solutions~$w$~of the field equations (67). Specifically we make the following ansatz 
\begin{equation}w=\sum_{p=0}^{q}\frac{Z^{p}}{p!}w^{(p)}~~~~\textrm{with}~~w(0)=w^{(0)}=u^{0}\end{equation}
and now note that
\begin{displaymath}Z(A^{0}(w)\partial_{z}w-A^{i}(w)\partial_{i}w-B(w)w)-C(w)w=\end{displaymath}
\begin{equation}\sum_{p=0}^{q}\frac{Z^{p}}{p!}(pA^{0}(w)-C(w))w^{(p)}-Z(A^{i}(w)\partial_{i}w+B(w)w)\end{equation}
Let the~$w^{(p)}$~be determined by demanding that the~$Z$-derivatives up to order~$q$~of the right hand side of (90) vanish at Z=0. To see that this prescription is well-defined note for example that the first derivative is
\begin{displaymath}\sum_{p=1}^{q}\frac{Z^{p-1}}{(p-1)!}(pA^{0}(w)-C(w))w^{(p)}+\sum_{p=0}^{q}\frac{Z^{p}}{p!}(pA^{0}(w)-C(w))'w^{(p)}\end{displaymath}
\begin{displaymath}-Z(A^{i}(w)\partial_{i}w+B(w)w)'-(A^{i}(w)\partial_{i}w+B(w)w)\end{displaymath}
Setting~$Z=0$~and equating the resulting expression to zero gives
\begin{displaymath}(A^{0}(u^{0})-C(u^{0}))w^{(1)}-(C(w))'(0)u^{0}-(A^{i}(u^{0})\partial_{i}u^{0}+B(u^{0})u^{0})=0\end{displaymath}
Lemma 4.1 now implies that~$w^{(1)}$~is uniquely determined by the initial data~$u^{0}$. Equating the pth derivative of (90) to zero at~$Z=0$~gives an equation of the form
\begin{displaymath}(pA^{0}(u^{0})-C(u^{0}))w^{(p)}+\phi(w^{(p-1)},\ldots)=0\end{displaymath}
Thus the~$w^{(p)}$~are inductively well-defined and are~$C_{0}^{\infty}$~by the choice of initial data.\\
By Taylor's theorem we must now have that
\begin{equation}Z(A^{0}(w)\partial_{z}w-A^{i}(w)\partial_{i}w-B(w)w)-C(w)w=\zeta\end{equation}
for some~$\zeta\in C_{0}^{\infty}$~admitting an estimate~$\|\zeta\|_{s}\leq\kappa_{s}Z^{q+1}$, say. It follows that
\begin{equation}A^{0}(w)\partial_{z}w-A^{i}(w)\partial_{i}w-B(w)w-\frac{1}{Z}C(w)w=\xi\end{equation}
where~$\xi\in C_{0}^{\infty}$~for each~$Z$~and admits an estimate~$\|\xi\|_{s}\leq\kappa_{s}Z^q$.\\
We now wish to check that the approximate solution~$w=(f_{q},\bar{g}_{q},\bar{g}^{-1}_{q},\partial\bar{g}_{q},\bar{Z}_{q})$\linebreak[4] of equations (67) gives rise to an approximate solution of the original localised reduced equations (56)-(57). First a calculation shows that the process of writing
\begin{displaymath}Z^{-1}(H_{i}(Z)-H_{i}(0))=\int_{0}^{1}H_{i}'(SZ)dS\end{displaymath}
which led to (67) can be inverted to obtain an approximate solution of equations (50) and (52) with error~$O(Z^{q})$. Now one needs to check that the part~$Z_{\alpha\beta}$~of the approximate solution is close to the original definition (51). To do this define~$\hat{Z}_{\mu\nu}$~by the right hand side of (51) expressed in terms of the appropriate parts of our existing approximate solution. Then one calculates that
\begin{equation}\partial_{z}(\hat{Z}_{\mu\nu}-Z_{\mu\nu}(0))=-\frac{1}{Z}(\hat{Z}_{\mu\nu}-Z_{\mu\nu}(0))+\frac{2}{Z}\{\star+O(Z^{q})\}\end{equation}
where~$\star$~is most of the right hand side of (60):
\begin{equation}\partial_{z}(Z_{\mu\nu}-Z_{\mu\nu}(0))=-\frac{1}{Z}(Z_{\mu\nu}-Z_{\mu\nu}(0))+\frac{2}{Z}\{\star\}+O(Z^{q})\end{equation}
Subtracting (93) from (94) gives
\begin{equation}\partial_{z}(Z(\hat{Z}_{\mu\nu}-Z_{\mu\nu}))=O(Z^{q})\end{equation} 
Hence~$\hat{Z}_{\mu\nu}-Z_{\mu\nu}=O(Z^{q})$.
\subsubsection{The reduced EV equations as a coupled symmetric hyperbolic system}
While the form (61)-(65), (67) of the reduced equations is useful for demonstrating formal solvability, it is not appropriate for the application of the method of energy estimates. The presence of the~$\frac{\partial f}{\partial x^{i}}$~terms in the equation for~$Z_{\mu\nu}$~spoils the symmetry of the system as a whole. On the other hand, while the field equations in the form (49)-(50) are symmetric, the~$\frac{1}{Z^{2}}$~terms are too strong to be handled by the contraction mapping technique of Claudel and Newman. We need to choose variables in such a way that the field equations are symmetric as a whole and have singular terms which are no worse than~$\frac{1}{Z}$. This can be achieved by defining the following new variable: 
\begin{equation}P_{\alpha\beta}=\phi(4\nabla_{\alpha}Z\nabla_{\beta}Z-g^{00}g_{\alpha\beta})-\frac{(\textrm{det}g)^{-1/2}}{p^{0}}fp_{\alpha}p_{\beta}\end{equation}
where~$\phi=\phi(p_{i})\in C_{0}^{\infty}$~and~$\int_{\mathbb{R}^{3}}\phi~d^{3}p=1$. We also suppose that~$\textrm{supp}\phi\subset\textrm{supp}f^{0}(x,\cdot)$. Now define
\begin{equation}Q_{\alpha\beta}=\frac{2}{Z}(P_{\alpha\beta}-P_{\alpha\beta}(0))\end{equation}
with~$Q_{\alpha\beta}(0)=2\partial_{z}P_{\alpha\beta}(0)$~determined by the localised data. In particular~$Q_{\mu\nu}(0)\in C_{0}^{\infty}(O_{\alpha}\times\mathbb{R}^{3})$~and~supp$Q_{\mu\nu}(0,x^{i},\cdot)\subset\textrm{supp}f^{0}(x^{i},\cdot)$.

The reduced field equations now decompose as
\begin{equation}\partial_{z}\bar{g}_{\alpha\beta}=h_{\alpha\beta 0}\end{equation}
\begin{equation}-g^{ij}\partial_{z}h_{\alpha\beta j}=-g^{ij}\partial_{j}h_{\alpha\beta 0}\end{equation}
\begin{equation}\partial_{z}\bar{g}^{\alpha\beta}=-g^{\alpha\mu}g^{\beta\nu}h_{\mu\nu 0}\end{equation}
\begin{displaymath}g^{00}\partial_{z}h_{\alpha\beta 0}=-g^{ij}\partial_{j}h_{\alpha\beta i}-2g^{0i}\partial_{i}h_{\alpha\beta 0}+F_{\alpha\beta}(g,h)\end{displaymath}
\begin{equation}+\frac{F(x^{i})}{Z}\left\{2g^{0\gamma}(h_{\alpha\gamma\beta}+h_{\beta\gamma\alpha}-h_{\alpha\beta\gamma})+\int_{\mathbb{R}^{3}}Q_{\alpha\beta}~d^{3}p\right\}\end{equation}
\begin{equation}\frac{\partial f}{\partial Z}=-\frac{p^{i}}{p^{0}}\frac{\partial f}{\partial x^{i}}+\frac{1}{2}(\partial_{i}g^{\alpha\beta})\frac{p_{\alpha}p_{\beta}}{p^{0}}\frac{\partial f}{\partial p_{i}}\end{equation}

Using the Vlasov equation one calculates the following evolution for~$Q_{\alpha\beta}$
\begin{displaymath}\partial_{z}Q_{\alpha\beta}=-\frac{p^{i}}{p^{0}}\frac{\partial Q_{\alpha\beta}}{\partial x^{i}}+\frac{1}{2}(\partial_{i}g^{\mu\nu})\frac{p_{\mu}p_{\nu}}{p^{0}}\frac{\partial Q_{\alpha\beta}}{\partial p_{i}}\end{displaymath}
\begin{displaymath}+\frac{1}{Z}\Bigg\{-Q_{\alpha\beta}+2\phi(g_{\alpha\beta}g^{0\mu}g^{0\nu}h_{\mu\nu 0}-g^{00}h_{\alpha\beta 0}-(p^{i}/p^{0})\partial_{i}(g^{00}g_{\alpha\beta}))\end{displaymath}
\begin{displaymath}f\Bigg(-2\partial_{z}\left(\frac{p_{\alpha}p_{\beta}}{p^{0}}\textrm{det}g^{-1/2}\right)-2\frac{p^{i}}{p^{0}}\partial_{i}\left(\frac{p_{\alpha}p_{\beta}}{p^{0}}\textrm{det}g^{-1/2}\right)\end{displaymath}
\begin{displaymath}+(\partial_{i}g^{\mu\nu})p_{\mu}p_{\nu}\frac{\partial~}{\partial p_{i}}\left(\frac{p_{\alpha}p_{\beta}}{p^{0}}\right)\textrm{det}g^{-1/2}\Bigg)+\frac{p_{\mu}p_{\nu}}{p^{0}}(\partial_{i}g^{\mu\nu})\frac{\partial\phi}{\partial p_{i}}(\bar{g}^{00}g_{\alpha\beta}+\bar{g}_{\alpha\beta})\end{displaymath}
\begin{equation}-2\frac{p^{i}}{p^{0}}\partial_{i}P_{\alpha\beta}(0)-\frac{p_{\mu}p_{\nu}}{p^{0}}(\partial_{i}g^{\mu\nu})\frac{\partial~}{\partial p_{i}}\left(\frac{fp_{\alpha}p_{\beta}}{p^{0}}\textrm{det}g^{-1/2}\right)(0)\Bigg\}\end{equation}
Clearly the system (98)-(103) is symmetric hyperbolic.

We now note that the terms containing~$p^{0}$~in equations (102)-(103) blow up at the vertex of the light cone and thus cause technical problems there. For this reason we multiply the right hand sides of these equations by a function~$\psi(p_{i})\in C_{0}^{\infty}(\mathbb{R}^{3})$~which is equal to unity on a set containing the support of~$f^{0}$~and equal to zero in a neighbourhood of the origin. Call the resulting equations $(102)',~(103)'$.

We now wish to show that the approximate solution~$(f_{q},g_{q},h_{q})$~of the field equations obtained earlier gives rise to an approximate solution\linebreak[4]~$(f_{q},g_{q},h_{q},Q_{q})$~of (98)-(101), $(102)'-(103)'$. To this end define~$Q_{\alpha\beta}=2Z^{-1}(P_{\alpha\beta}-P_{\alpha\beta}(0))$~with~$P_{\alpha\beta}$~constructed from~$f_{q},g_{q}$~via equation (96). If~$(f_{q},g_{q},h_{q})$~satisfies (56)-(57) up to~$O(Z^{q})$~then one has (dropping the subscript~$q$~for the moment)
\begin{displaymath}g^{00}\partial_{z}h_{\alpha\beta 0}=-g^{ij}\partial_{i}h_{\alpha\beta j}-2g^{0i}\partial_{i}h_{\alpha\beta 0}+F_{\alpha\beta}\end{displaymath}
\begin{equation}\frac{F(x)}{Z}\left\{\int_{\mathbb{R}^{3}}Q_{\alpha\beta}~d^{3}p+2g^{0\gamma}(h_{\alpha\gamma\beta}+h_{\beta\gamma\alpha}-h_{\alpha\beta\gamma})\right\}+O(Z^{q})\end{equation}
A calculation gives that this~$Q_{\alpha\beta}$~satifies
\begin{equation}\partial_{z}Q_{\alpha\beta}= rhs+O(Z^{q-1})\end{equation}
where~`$rhs$'~is the right hand side of equation $(141)'$. It follows that we have an approximate solution~$(f_{q-1},g_{q-1},h_{q-1},Q_{q-1})$~of the field equations with error~$O(Z^{q-1})$.
\subsubsection{The field equations in matrix form}
Let the vector~$u=u(Z,x^{i})$~stand for~$(\bar{g}_{\alpha\beta},\bar{g}^{\alpha\beta},h_{\alpha\beta\gamma})$. Then we may write (98)-(101) in the following form
\begin{equation}A^{0}(u)\frac{\partial u}{\partial Z}=\sum_{i=1}^{3}A^{i}(u)\frac{\partial u}{\partial x^{i}}+G(u)+\frac{F(x^{i})}{Z}\left\{C(u)+\int_{\mathbb{R}^{3}}\left(\begin{array}{l}0\\Q\end{array}\right)~d^{3}p\right\}\end{equation}
where~$A^{\alpha}, G, C$~are polynomial in the components of~$u$,~$A^{\alpha}$~are symmetric matrices and~$A^{0}$~is positive definite.

Let~$v=v(Z,x^{i},p_{j})$~stand for~$(f,Q_{\alpha\beta})$. Then we may write $(102)'-(103)'$ as
\begin{equation}\frac{\partial v}{\partial Z}=\sum_{i=1}^{6}B^{i}(p_{j},u)\partial_{i}v+\frac{1}{Z}D(u,v,p_{j})\end{equation}
where the~$B^{i}$~are symmetric matrices and~$B^{i}, D$~depend smoothly on~$p_{j}, v$~and on~$u$~away from det$g=0$.

Now let~$u_{q}$~and~$v_{q}$~stand for the appropriate parts of the approximate solution we obtained earlier. Then we have
\begin{displaymath}A^{0}(u_{q})\frac{\partial u_{q}}{\partial Z}=\sum_{i=1}^{3}A^{i}(u_{q})\frac{\partial u_{q}}{\partial x^{i}}+G(u_{q})+\frac{F(x^{i}}{Z}\left\{C(u_{q})+\int_{\mathbb{R}^{3}}\left(\begin{array}{l}0\\Q_{q}\end{array}\right)~d^{3}p\right\}\end{displaymath}
\begin{equation}+\xi_{1}\end{equation}
\begin{equation}\frac{\partial v_{q}}{\partial Z}=\sum_{i=1}^{6}B^{i}(p_{j},u_{q})\partial_{i}v_{q}+\frac{1}{Z}D(u_{q},v_{q},p_{j})+\xi_{2}\end{equation}
where~$\xi_{1}\in C_{0}^{\infty}(\mathbb{R}^{3})$~and~$\xi_{2}\in C_{0}^{\infty}(\mathbb{R}^{6})$~are~$O(Z^{q})$.

By considering the differences between (106), (108) and (107), (109) one is lead to the following linear system of equations
\begin{displaymath}A^{0}(U)\partial_{z}(u-u_{q})=A^{i}(U)\partial_{i}(u-u_{q})-(A^{0}(U)-A^{0}(u_{q}))\partial_{z}u_{q}\end{displaymath}
\begin{displaymath}+(A^{i}(U)-A^{i}(u_{q}))\partial_{i}u_{q}+(G(U)-G(u_{q}))\end{displaymath}
\begin{equation}\frac{F(x^{j})}{Z}\left\{C(U)-C(u_{q})+\int_{\mathbb{R}^{3}}\left(\begin{array}{cl}0\\Q(v)-Q_{q}\end{array}\right)~d^{3}p\right\}-\xi_{1}\end{equation}
\begin{displaymath}\partial_{z}(v-v_{q})=\sum_{i=1}^{6}B^{i}(p_{j},U)\partial_{i}(v-v_{q})+(B^{i}(p_{j},U)-B^{i}(p_{j},u_{q}))\partial_{i}v_{q}\end{displaymath}
\begin{equation}\frac{1}{Z}(D(U,V,p_{j})-D(u_{q},v_{q},p_{j}))-\xi_{2}\end{equation}
for the unknowns~$v-v_{q}, u-u_{q}$~with initial data~$v-v_{q}=u-u_{q}=0$. The quantities~$U$~and~$V$~are taken to be known smooth functions.~$Q_{q}$~stands for the appropriate part of~$v_{q}$.
\subsubsection{Solving the field equations by a contraction mapping technique}
By the theory of regular, linear symmetric hyperbolic PDE (Racke 1992) we can, given smooth~$(U,V)$~solve (111) for smooth~$v-v_{q}$~and then (110) for smooth~$u-u_{q}$~as long as~$U-u_{q}, V-v_{q}$~are~$O(Z^{q})$. These solutions exist as long as~det$g$~remains positive, where the elements~$g_{\alpha\beta}$~are extracted from the vector~$U$. Equations (110)-(111) thus generate a map~$\Phi:(U,V)\rightarrow (u,v)$. We will define a metric space~$(d,S)$~of functions in such a way that~$\Phi$~is a contraction mapping on~$S$~with respect to~$d$. Some standard techniques from the theory of regular quasilinear symmetric hyperbolic systems (Racke 1992) are used to achieve this.

Let~$U$~defined on~$\mathbb{R}^{3}$~and~$V$~defined on~$\mathbb{R}^{6}$~belong to the space~$S$~iff the following hold:
\begin{enumerate}
\item~$V(Z,x^{i},\cdot)$~is supported in~$\Omega$~for~$Z\leq T$, where~$\Omega$,~relatively compact, is slightly larger than supp$f^{0}(x^{i},\cdot)$~and supported outside~$B_{\epsilon}(0)$~for$Z\leq T$~with~$B_{\epsilon}(0)$~slightly smaller than supp$f^{0}$.
\item~The elements~$g_{\alpha\beta}$~of~$U$~are such that~$|\textrm{det}g(Z,x^{i})|\geq\delta~~\forall Z\leq T$.
\item~$U\in C_{0}^{\infty}(O_{\alpha}),~V\in C_{0}^{\infty}(O_{\alpha}\times\Omega)$.
\item max$\{\|U-u_{q}\|_{s},~\|V-v_{q}\|_{s}\}\leq \rho Z^{q}$~for~$Z\leq T,~\rho$~const,~$s$~large.
\item~$\|\partial_{z}U\|_{s-1}\leq L$
\end{enumerate}
Here~$\|~\|_{s}$~is the~$L^{2}$~type Sobolev norm of order~$s$~(Adams 1975). We choose~$s$~so large that the Sobolev imbedding theorem may be applied wherever needed in the sequel.

\textbf{Lemma 4.2} If~$T$~is chosen sufficiently small and~$L$~sufficiently large then~$\Phi:S\rightarrow S$.

~~~~\textit{Proof}~First we deal with condition 4. Apply~$x^{i}$~derivatives of order~$\alpha$~to (110) and~$x^{i}, p_{j}$~derivatives of order~$\alpha'$~to (111) to get, after some rearrangement 
\begin{equation}A^{0}(U)\partial_{z}\nabla^{\alpha}(u-u_{q})=A^{i}(U)\partial_{i}\nabla^{\alpha}(u-u_{q})+F_{1}^{\alpha}-\nabla^{\alpha}\xi_{1}\end{equation}
\begin{equation}\partial_{z}\nabla^{\alpha'}(v-v_{q})=B^{i}(x)\partial_{i}\nabla^{\alpha'}(v-v_{q})+F_{2}^{\alpha'}-\nabla^{\alpha'}\xi_{2}\end{equation}
where the~$F_{i}$~contain all the singular~$\frac{1}{Z}$~terms as well as terms like
\begin{displaymath}-A^{0}(U)\nabla^{\alpha}(A^{0}(U)^{-1}A^{j}(U)\partial_{j}(u-u_{q}))\end{displaymath}
etc.

Now take the inner product of (112) with~$\nabla^{\alpha}(u-u_{q})$~and of (113) with~$\nabla^{\alpha'}(v-v_{q})$. Using the symmetry of~$A^{\alpha}, B^{i}$~and the uniform equivalence of the~$L^{2}$~norm of~$\nabla^{\alpha}(u-u_{q})$~to
\begin{displaymath}\int A^{0}(U)\nabla^{\alpha}(u-u_{q})\cdot\nabla^{\alpha}(u-u_{q})~d^{3}x^{i}\end{displaymath}
in a standard way (Racke 1992) we sum derivatives of order~$\leq s$~to obtain
\begin{equation}\|v-v_{q}\|_{s}(Z)\leq c\int_{0}^{Z}\Big(\|v-v_{q}\|_{s}+\sum_{|\alpha|\leq s}\|F_{2}^{\alpha}\|_{2}\Big)(t)+t^{q}~dt\end{equation}
\begin{equation}\|u-u_{q}\|_{s}(Z)\leq c\int_{0}^{Z}\Big(\|u-u_{q}\|_{s}+\sum_{|\alpha'|\leq s}\|F_{1}^{\alpha'}\|_{2}\Big)(t)+t^{q}~dt\end{equation}
where we also used~$\|\xi_{i}\|_{s}\leq cZ^{q}$.

For the terms in~$F_{2}^{\alpha}$~coming from~$B^{i}$~one may readily use a Moser type estimate (Racke 1992) to obtain an~$L^{2}$~bound in terms of the~$H^{s}$~norm of~$v-v_{q}, U-u_{q}$. For the terms coming from~$D$~it is possible to obtain an~$L^{2}$~bound in terms of the~$H^{s}$~norm of~$U-u_{q}, V-v_{q}$~multiplied by~$\frac{1}{Z}$. In this way (114) leads to
\begin{equation}\|v-v_{q}\|_{s}\leq c\int_{0}^{Z}\|v-v_{q}\|_{s}+\|U-u_{q}\|_{s}+\frac{1}{t}(\|U-u_{q}\|_{s}+\|V-v_{q}\|_{s})+t^{q}~dt\end{equation}
Similarly the terms in~$F_{1}^{\alpha}$~coming from~$A^{\alpha}, G$~may be bounded in a standard way by~$\|u-u_{q}\|_{s}, \|U-u_{q}\|_{s}$. Terms coming from~$C$~can, by Moser and Sobolev inequalities be bounded by a constant times~$Z^{-1}\|U-u_{q}\|_{s}$. The bound (iv) on~$U$~gives a bound for the speed of propagation in equation (111) and thus a bound on the support of~$Q(v)$~in the~$p_{i}$~variables. This entails that the momentum integral of~$Q$~may be estimated by~$\|v-v_{q}\|_{s}$~, via the Cauchy-Schwarz inequality. In this way (115) leads to
\begin{equation}\|u-u_{q}\|_{s}\leq c\int_{0}^{Z}\|u-u_{q}\|_{s}+\|U-u_{q}\|_{s}+\frac{1}{t}(\|v-v_{q}\|_{s}+\|U-u_{q}\|_{s})+t^{q}~dt\end{equation}
Now write~$X=\|v-v_{q}\|_{s}+\|u-u_{q}\|_{s}$~and~$Y=\|U-u_{q}\|_{s}+\|V-v_{q}\|_{s}$. Adding (116) and (117) gives
\begin{equation}X(Z)\leq C\int_{0}^{Z}X+Y+t^{q}+\frac{1}{t}(X+Y)~dt\end{equation}
From (111) we see that
\begin{displaymath}\frac{\partial^{n}(v-v_{q})}{\partial Z^{n}}\Bigg\vert_{Z=0}=0\end{displaymath}
for~$n<q$. Thus~$v-v_{q}$~is~$O(Z^{q})$. Similarly~$u-u_{q}$~is~$O(Z^{q})$.\\Now put
\begin{equation}X'(Z)\equiv Z^{-q}X(Z)\qquad\|X\|'_{s}\equiv\sup_{0\leq t\leq T}t^{-q}\|X\|_{s}(t)\end{equation}
with similar definitions for~$Y$. Equation (118) now implies
\begin{equation}X'(Z)\leq C\left(Z^{-q}\int_{0}^{Z}X'(t^{q}+t^{q-1}+Y'(t^{q}+t^{q-1})~dt~+~Z\right)\end{equation}
and thus
\begin{equation}\|X\|'_{s}\leq C\left\{\left(\frac{T}{q+1}+\frac{1}{q}\right)\|X\|'_{s}+\left(\frac{T}{q+1}+\frac{1}{q}\right)\rho+T\right\}\end{equation}
By choosing~$q$~large and~$T$~small we can therefore arrange that~$\|X\|'_{s}\leq\rho$. It follows that~$\Phi$~preserves condition 4. Conditions 1, 2 and 3 follow from standard theory for linear hyperbolic PDE and from the finite speed of propagation inherent in the equations. 

Now given condition 4 for~$(U,V)$~it is elementary, using a Moser estimate to show that
\begin{equation}\|\partial_{z}(u-u_{q})\|_{s-1}\leq C_{1}\left(\|u-u_{q}\|_{s}+(1+Z^{-1})\|v-v_{q}\|_{s}\right)+C_{2}\end{equation}
But by the preceeding result the right hand side can be bounded by a constant. Thus if~$L$~is chosen large enough then~$\Phi$~preserves condition 5. Hence~$\Phi:S\rightarrow S$.

Now define a distance function~$d$~on~$S$~according to
\begin{equation}d(p,p')=\sup_{0\leq t\leq T}t^{-q}\|p-p'\|_{2}\end{equation}

\textbf{Lemma 4.3} For small~$T$~and large~$q$~the mapping~$\Phi$~is a contraction on~$S$~with respect to~$d$.

~~~~\textit{Proof}~Let~$(U,V),~(U',Y')\in S$~and put~$(u,v)=\Phi(U,V),~(u',v')=\Phi(U',V')$. Also write~$\Delta U=U-U'$~etc. Then by a standard energy argument one gets
\begin{equation}\|\Delta v\|_{2}\leq C\int_{0}^{Z}\|\Delta v\|_{2}+\|\Delta U\|_{2}+\frac{1}{t}(\|\Delta U\|_{2}+\|\Delta V\|_{2})~dt\end{equation}
and
\begin{equation}\|\Delta u\|_{2}\leq c\int_{0}^{Z}\|\Delta u\|_{2}+\|\Delta U\|_{2}+\frac{1}{t}\|\Delta v\|_{2}~dt\end{equation}
Now put~$X=\|\Delta v\|_{2}+\|\Delta u\|_{2},~Y=\|\Delta U\|_{2}+\|\Delta V\|_{2}$. Then
\begin{equation}X\leq C\int_{0}^{Z}X+Y+\frac{1}{t}(X+Y)~dt\end{equation}
If we now define
\begin{equation}\|X\|'_{2}=\sup_{0\leq t\leq T}t^{-q}\|u\|_{2}\end{equation}
with a similar definition for~$Y$~then (126) leads to
\begin{equation}\|X\|'_{2}\leq C\left(\frac{Z}{q+1}+\frac{1}{q}\right)\|X\|'_{2}+C\left(\frac{Z}{q+1}+\frac{1}{q}\right)\|Y\|'_{2}\end{equation}
For small~$T$~and large~$q$~one may thus arrange that~$\|X\|'_{2}\leq (1-\delta)\|Y\|'_{2}$~for some~$\delta > 0$. Hence the Lemma.

Now it is elementary to show that the set~$S^{\dagger}\supset S$~of functions~$p$~satisfying the condition
\begin{displaymath}\sup_{0\leq t\leq T}t^{-q}\|p\|_{2}\leq\rho\end{displaymath}
is complete with respect to the metric~$d$. Hence there exists~$p\in\bar{S}$~such that~$p=\Phi(p)$. Now let~$p^{k}=\Phi^{k}(p^{0})$~for some~$p^{0}\in S$. Then~$d(p^{k},p)\rightarrow 0$~as~$k\rightarrow\infty$~and hence~$p^{k}$~is Cauchy with respect to~$d$.

By the Gagliardo-Nirenberg inequality (Racke 1992) one now obtains
\begin{equation}\|Z^{-q}(p^{k}-p^{m})\|_{s'}\leq \|Z^{-q}(p^{k}-p^{m})\|_{s}^{s'/s}\|Z^{-q}(p^{k}-p^{m})\|_{2}^{1-s'/s}\end{equation}
for~$s'<s$.\\
But~$\Phi:S\rightarrow S$~and so the~$\|Z^{-q}p^{k}\|_{s}$~are uniformly bounded. It follows that if we define the distance function~$d_{s'}$~by
\begin{displaymath}d_{s'}(p,p')=\sup_{0\leq t\leq T}t^{-q}\|p-p'\|_{s'}\end{displaymath}
then~$p^{k}$~is Cauchy with respect to~$d_{s'}$. Hence~$\|p^{k}-p\|'_{s'}\rightarrow 0$~as~$k\rightarrow\infty$~and~$\|p\|'_{s'}\leq\rho$.\\
From the field equations one sees that the~$\partial_{z}p^{k}$~converge in~$C([0,T], H^{s'-1})$\linebreak[4]and therefore~$p\in C([0,T],H^{s'})\cap C^{1}([0,T],H^{s'-1})$. By the Sobolev imbedding theorem~$p$~is classically~$C^{1}$~and satifies the field equations (~$p$~stands for the whole solution here). Since~$p$~is~$O(Z^{q})$~we can repeatedly differentiate the field equation to get~$p\in\bigcap_{p=0}^{s'}C^{p}([0,T],H^{s'-p})$. It follows that the parts of~$p$~corresponding to~$f, g_{\alpha\beta}$~and~$h_{\alpha\beta\gamma}$~belong to~$\bigcap_{p=0}^{s'}C^{p}([0,T],H^{s'-p})$. We could of course have chosen~$s'$~arbitrarily large and we thus get that~$f, g_{\alpha\beta}$~and\linebreak[4]$h_{\alpha\beta\gamma}$~belong to~$C([0,T_{s}],H^{s})$, for~$T_{s}\leq T, s>0$. The data at~$T_{s}$~can then be used to construct data for the regular Einstein-Vlasov equations. Now it is well-known that the evolution of~$H^{s}$~data for the regular EV equations stays in~$C([0,Z],H^{s})$~as long as the~$C^{1}$~norm remains bounded (Rendall 1997). In this way we see that our solution of the conformal field equations lies in~$C([0,T],H^{s})$~for every $s$~and is therefore~$C^{\infty}$.

It remains to show that the part~$Q_{\alpha\beta}$~of our solution agrees with the original definition (96)-(97). If we use the parts~$f$~and~$g_{\alpha\beta}$~of our solution to construct a quantity~$\hat{Q}_{\alpha\beta}$~via (96)-(97) then a calculation gives that
\begin{equation}\partial_{z}(\hat{Q}_{\alpha\beta}-Q_{\alpha\beta})=\sum_{i=1}^{6}M^{i}(u, p_{j})\partial_{i}(\hat{Q}_{\alpha\beta}-Q_{\alpha\beta})-\frac{1}{Z}(\hat{Q}_{\alpha\beta}-Q_{\alpha\beta})\end{equation}
where the~$M^{i}$~are symmetric.

Since the singular part of this equation is `negative definite' an~$L^{2}$~energy estimate gives that~$Q_{\alpha\beta}=\hat{Q}_{\alpha\beta}$~as required.

Also note that for small~$T'$~we have~$F(x)=\Phi(p)=1$~on the domain of dependence of~$O'_{\alpha}\times [0,T']$. Thus~$u$~solves the conformal Einstein-Vlasov equations on~$O'_{\alpha}\times [0,T']$~and is determined by the right data.

\subsubsection{Uniqueness of solutions}
We now show that the solution~$p$~constructed above is the only smooth solution of the conformal field equations with the given data.

Suppose then that we have two solutions~$(f, g_{\alpha\beta}),~(\hat{f}, \hat{g}_{\alpha\beta})$~of the localised equations (49)-(50) with the same~$C_{0}^{\infty}$~data. These two solutions will then satisfy equations (96)-(97), (106)-(107). Let~$u$~stand for~$(g, \partial g)$~and let~$v$~stand for~$(f, Q)$~with similar definitions for~$\hat{u},~\hat{v}$. Since the system (106)-(107) is symmetric hyperbolic and the supports of~$f$~and~$\hat{f}$~remain bounded on any closed time interval of existence we may obtain the following~$L^{2}$~energy estimates for the differences~$\Delta u$~and~$\Delta v$:
\begin{equation}\|\Delta u\|_{2}(Z)\leq C\int_{0}^{Z}\|\Delta u\|_{2}+\frac{1}{s}(\|\Delta u\|_{2}+\|\Delta v\|_{2})~ds\end{equation}
\begin{equation}\|\Delta v\|_{2}(Z)\leq C\int_{0}^{Z}\|\Delta u\|_{2}+\|\Delta v\|_{2}+\frac{1}{s}(\|\Delta u\|_{2}+\|\Delta v\|_{2})~ds\end{equation}
Now letting~$X=\|\Delta u\|_{2}+\|\Delta v\|_{2}$~we get that
\begin{equation}X(Z)\leq C\int_{0}^{Z}(1+s^{-1})X~ds\end{equation}
We know from section 4.4.2 that the Taylor series of a solution to the field equations is determined by the initial data. Thus our two solutions~$(u, v)$~and\linebreak[4]$(\hat{u}, \hat{v})$~must agree up to arbitrary order in~$Z$. This is to say that~$X\leq\lambda_{q}Z^{q}$~for arbitrary~$q$. Plugging this into (133) gives
\begin{equation}X(Z)\leq C\lambda_{q}Z^{q}\left(\frac{1}{q}+\frac{Z}{q+1}\right)\end{equation}
Now choose~$q>C$~and choose~$T$~small enough so that the right hand side of (134) is less than~$(1-\epsilon)\lambda_{q}Z^{q}$~on~$[0,T]$~for some~$\epsilon >0$. We see iteratively that~$X\leq (1-\epsilon)^{n}\lambda_{q}Z^{q}$~on~$[0,T]$. Thus~$\Delta u=\Delta v=0$~as required.
\subsubsection{Dependence of solutions on initial data}
Suppose now that we have a 1-parameter family of small perturbations\linebreak[4]$f^{0}_{\epsilon}(x,p)$~of some initial datum~$f^{0}_{0}$~depending smoothly on~$\epsilon, x$~and~$p$~for~$\epsilon\in(-1,1)$. We wish to show that the corresponding solutions~$(f_{\epsilon}, g^{\epsilon}_{ij})$~of the conformal field equations depend smoothly on the parameter~$\epsilon$. 
First it follows from an analogue of Theorem 3.1 that the family of initial metrics~$h^{\epsilon}_{ij}$~depends smoothly on~$\epsilon$~and~$x^{i}$. Now we may reconsider the field equations (136)-(141) for~$f, g, h, Q$~with~$\epsilon$~simply acting as an extra coordinate. By replacing the spaces~$H^{s}(\mathbb{R}^{3})$~and~$H^{s}(\mathbb{R}^{6})$~respectively with\linebreak[4]$H^{s}(\mathbb{R}^{3}\times (-1,1))$~and~$H^{s}(\mathbb{R}^{6}\times (-1,1))$~we can now repeat the analysis of section 4.4.5 to get that the solution~$(f, g_{\alpha\beta})$~depends smoothly on~$\epsilon$.
\\\\\\
The results of section 4.4 may now be summarised as follows:\\\\

\textbf{Theorem 4.1} Given a smooth positive function~$f^{0}$~on the cotangent bundle of a paracompact 3-manifold~$M$~satisfying the integral constraint (13) and such that for~$x$~in a compact set
\begin{enumerate}
\item supp$f^{0}(x,\cdot)\subset\Omega_{1}$,~for some relatively compact~$\Omega_{1}\subset\mathbb{R}^{3}$
\item~$\mathbb{R}^{3}\setminus$supp$f^{0}(x,\cdot)\supset\Omega_{2}$, for some neighbourhood of the origin~$\Omega_{2}\subset\mathbb{R}^{3}$.
\end{enumerate}
there exists a unique solution~$(M\times(0,T), f,g_{ab})$~of the Einstein-Vlasov equations with an isotropic singularity having~$f^{0}$~as the orthogonal projection of the limiting particle distribution function at the singularity surface. The solution obtained depends smoothly on the choice of initial data.

\section*{Appendix}
\textbf{Theorem A.1.} Consider a PDE of the form
\begin{gather}A^{0}(u)\partial_{t}u=A^{i}(u)\partial_{i}u+B(u)u+\frac{1}{t}C(u)u\tag{1}\\
u(0)=u^{0}\in C_{0}^{\infty}(\mathbb{R}^{n})\tag{2}\end{gather}
with~$A^{\alpha}$~symmetric and~$A^{0}$~positive definite.\\
If~$C(u)u^{0}=0$~and~$(A^{0})^{-1}C(u^{0})$~has no positive integer eigenvalues
then there exists at most one smooth solution~$u$~of (1) with~$u(0)=u^{0}$.

\textit{Proof.}~Suppose we have two smooth solutions~$u,~v$~of (1) with~$u(0)=v(0)=u^{0}$. One may now subtract the equation satisfied by~$u$~from that satisfied by~$v$~and obtain the following energy estimate:
\begin{gather}\|u-v\|_{2}(t)\leq K\int_{0}^{t} \|u-v\|_{2}+\frac{1}{s}\|u-v\|_{2}~ds\tag{3}\end{gather}
on the time interval~$[0,T]$~for some constant~$K$.\\
Next note that the Taylor series of any smooth solution of (1), (2) is determined by~$u^{0}$, by the following argument:\\
Write~$u=u^{0}+tu^{1}+t^{2}u^{2}+\ldots +t^{q}R(t,x^{i})$. Then (1) implies
\begin{gather}A^{0}(u)\{u^{1}+2tu^{2}+\ldots +qt^{q-1}R+t^{q}R'\}\notag\end{gather}
\begin{gather}-A^{i}(u)\partial_{i}u-B(u)u-\{u^{1}+tu^{2}+t^{q-1}R\}=0\tag{4}\end{gather}
Evaluating (4) at~$t=0$~gives
\begin{gather}(A^{0}(u^{0})-C(u^{0}))u^{1}= A^{i}(u^{0})\partial_{i}u^{0}+B(u^{0})u^{0}\notag\end{gather}
and thus~$u^{1}$~is determined by~$u^{0}$.\\
Now we differentiate (4)~$p$~times with respect to time and evaluate at~$t=0$~to get an equation of the form
\begin{gather}(pA^{0}(u^{0})-C(u^{0}))u^{p}= F(u^{p-1},u^{p-2}\ldots)\notag\end{gather}
and thus~$u^{p}$~is determined inductively by~$u^{0}$.\\
Since our two solutions~$u, v$~of (1) have the same Taylor series it follows that we must have the inequality~$\|u-v\|_{2}\leq \lambda_{q}t^{q}$~for~$t\in [0,T]$~and~$q$~arbitrary. Substituting this into the right hand side of (3) leads to
\begin{gather}\|u-v\|_{2}\leq K\lambda_{q}t^{q}\left(\frac{1}{q}+\frac{t}{q+1}\right)\tag{5}\end{gather}
Now choose~$q> K$~and choose~$T$~so small that the right hand side of (5) is less than~$(1-\epsilon)\lambda_{q}t^{q}$~for some~$\epsilon>0$. Iteratively one gets~$\|u-v\|_{2}\leq (1-\epsilon)^{n}\lambda_{q}t^{q}$~on~$[0,T]$. Thus~$u=v$~as required.
\pagebreak
 
\section*{References}
\begin{description}

\item~K. Anguige, K. P. Tod, 1998, ~\emph{Isotropic Cosmological Singularities I}, \\gr-qc/9903008
\item~K. Anguige, K. P. Tod, 1998, ~\emph{Isotropic Cosmological Singularities II}, \\gr-qc/9903009
\item~C. M. Claudel, K. P. Newman, 1998~\emph{The Cauchy problem for quasilinear hyperbolic evolution problems with a singularity in the time, Proc. R. Soc. Lond}~454, 1073-1107
\item~R. A. Adams, 1975,~\emph{Sobolev spaces}~(New York: Academic Press)
\item~A. D. Rendall, 1997,~\emph{An introduction to the Einstein-Vlasov system}, in~\emph{Mathematics of Gravitation}, ed. P. Chrusciel (Banach Center Publications, Warszawa) vol. 41, part 1
\item~J. Ehlers, 1971,~in~\emph{General Relativity and cosmology}, ed. R. K. Sachs,~\emph{Varenna Summer School XLVII}~(Acad. Press N. Y)
\item~R. Racke, 1992,~\emph{Lectures on nonlinear evolution equations, Aspects of Mathematics}~vol. E19 (Vieweg)
\end{description}
\end{document}